\documentstyle[12pt]{article}

\advance\textwidth 1.5cm
\advance\oddsidemargin -0.75cm
\newcommand\beq{\begin{equation}}
\newcommand\eeq{\end{equation}}
\def\beqa{\begin{eqnarray}}
\def\eeqa{\end{eqnarray}}
\def\bega{\begin{array}}
\def\enda{\end{array}}

\def\non{{\nonumber}}

\def\vl{{\phantom{F}}}
\def\P{\protect}
\def\ft#1#2{{\textstyle{{#1}\over{#2}}}}
\def\im{{\rm i}}
\def\bl{{\biggl(}}
\def\br{{\biggr)}}

\def\H#1{{H^{(#1)}}}
\def\ps#1{{\psi^{(#1)}}}

\def\a{\alpha}
\def\b{\beta}

\def\G{\Gamma}

\def\d{\partial}

\begin{document}

\title{Special functions from quantum canonical transformations}
\author{Arlen Anderson\thanks{arley@ic.ac.uk}\\
Blackett Laboratory\\
Imperial College\\
Prince Consort Road\\
London SW7 2BZ England}
\date{Oct. 26, 1993, rev. May 27, 1994}
\maketitle

\vspace{-10cm}
\hfill Imperial-TP-93-94-5

\hfill hep-th/9310168
\vspace{10cm}

\begin{abstract}
Quantum canonical transformations are used to derive the integral
representations and Kummer solutions of the confluent hypergeometric and
hypergeometric equations.  Integral representations of the
solutions of the non-periodic three body Toda equation are also found.
The derivation of these representations  motivate the form of a
two-dimensional generalized hypergeometric equation which
contains the non-periodic Toda equation as a special case and
whose solutions may be obtained by quantum canonical transformation.
\end{abstract}

\section{Introduction}

The confluent hypergeometric and hypergeometric equations
underlie many of the special functions of mathematical physics and,
by extension, many of the exactly soluble problems of quantum
mechanics in one dimension.  The solutions of these equations have of
course been well understood for more than a hundred years (see e.g.
\cite{WhW}).  This paper returns to derive the integral representations
and Kummer solutions of the confluent
hypergeometric and hypergeometric equations using a new approach
whose motivation
comes from physics.  The purpose of this is two-fold: to give an elegant
and economical derivation of time-honored results and to illustrate
the use of some basic tools which have application to solving
more difficult problems.  As an illustration, simple derivations of two
integral representations of the wavefunctions for the (non-periodic)
three-body Toda equation are given.  This motivates a 2-dimensional
generalized hypergeometric equation which includes the non-periodic
Toda equation as a special case.

The origin of the method to be used lies in quantum mechanics.  Given a
linear differential equation, it can be interpreted as a time-independent
Schrodinger equation at a given energy, where the dependent variable is
taken to be the wavefunction and the differential operator to be the
Hamiltonian in the
coordinate representation. (For convenience, the discussion will be given
assuming the differential equation is in terms of a single variable; the
generalization to many variables is straightforward.)
The differential operator {\it cum}
Hamiltonian can be abstracted from the coordinate representation and
viewed directly as a function of non-commuting variables $x$ and $\d$
which satisfy the commutation relation $[x,\d]=-1$.  The coordinate
representation consists in taking $\d=d/dx$.  We use $\d$ instead of the
physicists' $p$ to avoid unnecessary factors of $i$; expressions in terms
of $p$ are simply obtained by substituting $\d=ip$.

The method consists in making a quantum canonical transformation from the
given Hamiltonian to another one having known solutions.  A quantum
canonical transformation is a change of variables $(x,\d)\mapsto (x'(x,\d),
\d'(x,\d))$ which preserves the algebraic relation $[x,\d]=-1=[x',\d']$.
It is so called by analogy to a classical canonical transformation which
is a change of the phase space variables which preserves the Poisson
bracket.  Quantum canonical transformation are produced by acting with
a function $C(x,\d)$,
\beq
q\mapsto q'=C q C^{-1},\qquad \d \mapsto \d'=C \d C^{-1}.
\eeq
Under this transformation, any function of $x$ and $\d$ transforms
\beq
\label{qct1}
H(x,\d)\mapsto H'(x,\d)=C H(x,\d) C^{-1}=H(C x C^{-1}, C\d C^{-1}).
\eeq
A solution $\psi$ of $H\psi=0$ is obtained from a solution $\psi'$ of
$CHC^{-1}\psi'=H'\psi'=0$ by
\beq
\psi=C^{-1}\psi'.
\eeq
Care must be taken if the kernel of $C$ as an operator in the coordinate
representation is non-trivial as not all $\psi$ obtained in this way
will be solutions of $H$ (see \cite{And}).

\section{Quantum canonical transformations}

To be more precise about the quantum canonical
transformations, one must address the nature of functions $C(x,\d)$
and their representation as operators.
By introducing $x^{-1}$ and $\d^{-1}$ as the algebraic inverses of $x$ and $\d$
($xx^{-1}=x^{-1}x=1$, $\d \d^{-1}=\d^{-1} \d=1$), one can define an
algebra $\cal U$ consisting of functions of the variables $x$
and $\d$ and their inverses, consistent with the commutation relation
$[x,\d]=1$.
The algebra is constructed so that every element $C\in {\cal U}$ has an
algebraic inverse $C^{-1}\in {\cal U}$, and so that elements transform
under a quantum canonical transformation by (\ref{qct1}).
Details are given in Ref. \cite{And}.
The precise class of functions has not been identified
with a named class, but
it essentially consists of those functions and their algebraic inverses
which do not involve distributions in themselves or any of their
derivatives (the functions may have poles, branch points or essential
singularities).  Every linear differential
operator with suitably smooth coefficients
corresponds to an element of this algebra.  Very likely it is possible to
generalize
the analysis to treat operators whose coefficients are not smooth,
but this has not been investigated.

The functions $C(x,\d)$ and their inverses are well-defined as elements
of the algebra ${\cal U}$.  Their realization as operators in the
coordinate representation is more subtle.  (For convenience, the same
notation will be used for the elements of the algebra and their realization
as operators.)  Care must be taken because the
operators corresponding to $C$ and $C^{-1}$ are not strictly inverses because
their kernels may be non-vanishing.  One can define them as inverses when
restricted to act on suitable
subspaces with the kernels projected out\cite{And}, but it is practical
to think of them as being inverses up to arbitrary linear combinations of
elements in their respective kernels.  For example,  the inverse of
differentiation, $\d=d/dx$, is indefinite integration, $\d^{-1}=\int^x dx$,
up to an arbitrary additive constant (an element of $\ker \d$).

One might be concerned about possible problems in defining the operator
realization of the quantum canonical transformations.
The property that makes the quantum canonical transformations useful
is that
they can be decomposed into a product of elementary transformations, each
of which has a well-defined action on wavefunctions\cite{And}.
There are three elementary canonical transformations:\newline
1) similarity transformations, $C=\exp(\int^x f(q) dq)$,
\beq
x\mapsto x,\quad \d\mapsto \d-f(x):\quad   C\psi(x)=\exp(\int^x f(q)
dq)\psi(x),
\eeq
2) point canonical transformations, $P_{f(x)}$,
\beq
x\mapsto f(x),\quad \d\mapsto {1\over f'(x)}\d: \quad
P_{f(x)}\psi(x)=\psi(f(x)),
\eeq
and 3) interchange, $I$,
\beq
x\mapsto -i \d,\quad \d\mapsto -i x:\quad I\psi(x)={1\over (2\pi)^{1/2}}
\int_{-\infty}^\infty dy e^{ix y}\psi(y).
\eeq
The functions in each transformation are allowed to be complex.
In the differential equation context, these correspond to similarity
transformations, change of variables, and Fourier transform.  By composing
them, one obtains ordered functions of operators whose behavior on
wavefunctions is well-defined.  In particular,
$$I\exp(\int^{i x} f(q) dq)I^{-1}=\exp(\int^{\d} f(q) dq)$$
and
$$IP_{f(i x)} I^{-1}=P_{f(\d)}$$
give the analogous similarity $(x,\d)\mapsto (x+f(\d),\d)$ and point
canonical transformations $(x,\d)\mapsto (f'(\d)^{-1}x,f(\d))$
involving $\d$.  These are the basic tools for making
quantum canonical transformations.  They encode in an
efficient way the details of performing the standard manipulations of
differential equations.

As an illustration, consider the transformation generated by $t^{x\d}$
which will be used later.  This can be decomposed as
\beq
t^{x\d}=\exp(\ln t\,x\d)=P_{\ln x} \exp(\ln t\,\d) P_{e^x}.
\eeq
The middle transformation is recognized as a translation while the
outer ones are point canonical transformations.
Their collective action on the algebraic element $x$ is then
\beqa
t^{x\d}x t^{-x\d} &=& P_{\ln x} \exp(\ln t\,\d) P_{e^x} x P_{\ln x}
\exp(-\ln t\,\d) P_{e^x}  \\
&=& P_{\ln x} \exp(x+\ln t) P_{e^x} \non \\
&=& t x, \non
\eeqa
and their action on a wavefunction is also to rescale $x$
\beq
\label{scale}
t^{x\d}\psi(x)=P_{\ln x} \exp(\ln t\,\d) P_{e^x}\psi(x)=\psi(tx).
\eeq
Alternatively, one could simply have observed that since
$$
x^{-1} t^{x\d}x = t^{x^{-1} x\d x}= t^{x\d+1},
$$
one has
$$
t^{x\d} x t^{-x\d}=x t^{x\d+1} t^{-x\d} = x t.
$$
For simple manipulations like this, it is often not necessary to explicitly
decompose the transformation into elementary canonical transformations,
though it may be done.

There is one main difference between the use of quantum canonical
transformations for physics and for simply solving differential
equations.  In physics, one often wants to find a transformation
which takes the entire spectrum of an operator into that of another.  In
particular, one would like to find a canonical transformation from an
interacting to a free theory.  If one is interested instead in an
equation like the hypergeometric equation, one is not concerned with the
spectrum of the operator, but rather with a particular solution.  The
significance of this is that one does not need to transform to an
operator for which one can find the general solution, only to one for which
a single solution is easily identified.  This will generate a single
solution of the operator one is interested in.  Other solutions can be
obtained from other canonical transformations, especially from
those which act as symmetries of one's
operator, taking it into another operator of the same form but with
different parameters.

\section{Confluent Hypergeometric Function}

The differential operator for the confluent hypergeometric equation\cite{AbS}
can be represented by the Hamiltonian
\beq
\label{chop}
H=x \d^2 +(b-x) \d -a.
\eeq
The confluent hypergeometric functions are the solutions $\psi$ of
\beq
H\psi=0.
\eeq
The Hamiltonian is easily canonically transformed to an equation soluble
on inspection after it is rewritten as
\beq
\label{CHeq}
H=(b+x \d ) \d -(a+x \d ).
\eeq

The quantum canonical transformation produced by
\beq
C={\G(b+x \d) \over \G(a+x \d)}.
\eeq
transforms the derivative
\beq
\d \mapsto C \d C^{-1}={a+x \d \over b+x \d} \d.
\eeq
This is easily verified by observing
$$ \d \G(a+x\d) \d^{-1} =\G(a+ \d x)=\G(a+x \d+1).$$
Since $C$ commutes with $b+x \d$ and $a+x \d$, it transforms the Hamiltonian to
\beq
\H{a}=CHC^{-1}=(a+x \d)(\d-1).
\eeq
One of the solutions of $\H{a}\ps{a}=0$ is clearly $\ps{a}=e^x$.  The
complementary solution is easily found but won't be needed.   (In a
problem for which one was interested in the complete spectrum $H\psi=E\psi$,
one wouldn't be satisfied with this form.  One would look for further
transformations leading to, say, a free Hamiltonian $H'=\d^2$.)

The solution of the confluent
hypergeometric equation which corresponds to $\ps{a}=e^x$ is
\beq
\psi=C^{-1} \ps{a}= {\G(a+x \d) \over \G(b+x \d)}e^{x}.
\eeq
Expanding the exponential in a power series in $x$ and allowing the
operators to act, one finds the familiar power series representation of
the confluent hypergeometric function
\beqa
\label{M}
\psi&=&\sum_{n=0}^\infty {\G(a+n) \over \G(b+n) } {x^n\over n!} \\
&=& {\G(a)\over \G(b)} M(a,b,x). \non
\eeqa

One can obtain an integral representation of (\ref{M}) by recognizing that
$C^{-1}$ times $\G(b-a)$ is a beta function,
\beq
{\G(a+x \d)\G(b-a) \over \G(b+x \d)} =\int_0^1 dt\, t^{a+x \d-1}
(1-t)^{b-a-1}
\eeq
($\Re(b)>\Re(a)>0$).
Applying this to $e^x$, and using (\ref{scale}), one finds
\beq
{\G(a)\G(b-a)\over \G(b)} M(a,b,x)=\int_0^1 dt\, t^{a-1} (1-t)^{b-a-1}
e^{tx}.
\eeq
This is a familiar integral representation\cite{AbS} from which others
may be obtained by change of variables.

A Barnes-type contour integral representation is also easily obtained by
expressing $e^x$ as
\beq
\label{Beq}
e^{x}={1\over 2\pi i} \int_{-i\infty}^{i\infty}ds\, \G(-s) (-x)^s .
\eeq
Applying $C^{-1}$ to this, one has
\beq
{\G(a)\over \G(b)} M(a,b,x)={1\over 2\pi i} \int_{-i\infty}^{i\infty}ds\,
{\G(-s)\G(a+s)\over \G(b+s)} (-x)^s.
\eeq
One must choose the contour of integration so that it separates the
poles of $\G(-s)$ and $\G(a+s)$.
Again, this is the standard result\cite{AbS}.

An expression for the other familiar confluent hypergeometric
function\cite{AbS}
$U(a,b,x)$ can be found from a second canonical transformation.  The
transformation
\beq
C'={1\over \G(a+x \d) \G(1-b-x \d)}
\eeq
transforms the momentum
\beq
\d \mapsto -{a+x \d\over b+x \d} \d.
\eeq
The transformed Hamiltonian is
\beq
\H{a'}=(a+x \d)(-\d-1).
\eeq
This has the immediate solution $e^{-x}$.
The confluent hypergeometric function corresponding to $e^{-x}$ is
\beq
\G(a)\G(1+a-b) U(a,b,x)=\G(a+x \d)\G(1-b-x \d) e^{-x} .
\eeq
One recognizes another instance of the beta function,
\beq
{\G(a+x \d)\G(1-b-x \d)\over \G(1+a-b)}=\int_0^\infty dt\,
    t^{a+x \d-1} (1+t)^{b-a-1}.
\eeq
Applying this to $e^{-x}$ gives
\beq
\G(a)U(a,b,x) =\int_0^\infty dt\, t^{a-1} (1+t)^{b-a-1} e^{-xt}.
\eeq

Additional forms of solutions can be found by applying symmetry
transformations of the equation.  These give the Kummer relations between
solutions.  A symmetry is a canonical transformation which takes an
operator into another operator of the same form but with different parameters.
Since we are interested here in equations of the form $H\psi=0$,
the term ``symmetry'' will be broadened to include transformations
which produce an operator of the form of $H$ up to an overall factor,
which can be divided out.
Let $H(a,b)$ denote the confluent hypergeometric operator (\ref{chop}).
A symmetry $C$ gives
\beq
CH(a,b)C^{-1}=FH(a',b')
\eeq
where $F$ is an overall factor.
Given a solution $\psi'$ of $H(a',b')$, $\psi=C^{-1}\psi'$ is a solution
of $H(a,b)$ (assuming here that the kernels of $F$ and $C$ vanish; a more
general discussion is possible\cite{And}).

The simplest symmetries are those obtained by similarity transformation.
Since the coefficient of the highest derivative term is $x$,
the transformation $C_1=x^{-\a}$, taking
\beq
\d \mapsto C_1\d C_1^{-1}=\d+{\alpha\over x},
\eeq
will produce a linear derivative term of correct form.
The transformed Hamiltonian is
\beq
\H{a}=x \d^2 +(2\a + b -x) \d +{\a^2-\a(1-b)\over x} -(a+\a).
\eeq
Taking $\a=0$ or $1-b$
cancels the undesired $1/x$ potential term,
and leaves an operator of confluent hypergeometric
form.  The symmetry is
\beq
C_1 H(a,b) C_1^{-1} =H(a+\a,b+2\a),
\eeq
where $\a=0,1-b$.  The transformation with $\a=0$ is trivial, but the
other produces a solution of $H(a,b)$
\beq
\psi=C_1^{-1}\ps{a}=x^{1-b}M(a-b+1,2-b,x).
\eeq
Since this solution has different behavior than (\ref{M}) at the regular
singular point $x=0$, it is an independent solution.

A second symmetry transformation is obtained from the transformation
$C_2=e^{-\b x}$,
\beq
\d\mapsto \d+ \beta.
\eeq
This leads to
\beq
\H{a}=x\d^2 +(b-(1-2\b)x)\d +b\b -a +(\b^2-\b)x.
\eeq
Requiring the coefficient of $x$ to vanish gives $\b=1$. The transformed
Hamiltonian is
\beq
\H{a}=x\d^2+(b+x)\d+b-a.
\eeq
This is not of confluent hypergeometric form because the sign of $x$ in the
linear term is wrong.
A second transformation $S=\exp(i\pi x \d)$
\beq
x\mapsto -x,\quad \d\mapsto -\d
\eeq
takes $\H{a}$ to $\H{b}$ in canonical form (\ref{chop}) up to an overall sign.
This sign is not important because the equation to be satisfied is
$H(a,b)\psi=0$ and an overall factor can be divided out.  The symmetry
here is
\beq
SC_2 H(a,b) C_2^{-1}S^{-1}=-H(b-a,b).
\eeq
A solution of (\ref{chop}) is then given by
\beq
\psi=C_2^{-1} S^{-1} \ps{b}=e^x M(b-a,b,-x).
\eeq

Combining these two transformations gives a third transformation,
\beq
C_1SC_2 H(a,b) C_2^{-1}S^{-1}C_1^{-1}= -H(b-a+\a,b+2\a),
\eeq
where $\a=1-b$.  This leads to the solution
\beq
\psi=x^{1-b} e^x M(1-a,2-b,-x).
\eeq
Inspection of the behavior near the regular singular point $x=0$ shows that
\beqa
M(a,b,x)&=& e^x M(b-a,b,-x), \\
x^{1-b}M(a-b+1,2-b,x) &=&  x^{1-b} e^x M(1-a,2-b,-x). \non
\eeqa
These are the Kummer transformations\cite{AbS}.  The above symmetry
transformations may also be applied to $U(a,b,x)$ to find similar
relations.

\section{Hypergeometric Equation}

A similar story can be repeated for the hypergeometric equation.  The
Hamiltonian corresponding to the hypergeometric equation is\cite{AbS2}
\beq
\label{hop}
H=x(1-x) \d^2 +(c-(a+b+1)x)\d -ab.
\eeq
This can be rewritten as
\beq
H=(c+x \d) \d -(a+x \d)(b+x \d).
\eeq
The canonical transformation
\beq
C={\G(c+x \d)\over \G(a+x \d)\G(b+x \d)}
\eeq
transforms the derivative
\beq
\d\mapsto C\d C^{-1}= {(a+x \d)(b+x \d)\over c+x \d} \d.
\eeq
The transformed Hamiltonian is
\beq
\H{a}=(a+x \d)(b+x \d)(\d-1).
\eeq
This has $\ps{a}=e^x$ as a solution.
The solution of $H$ corresponding to $\ps{a}$ is the familiar
hypergeometric function
\beqa
{\G(a)\G(b)\over \G(c)}F(a,b;c;x)&=&C^{-1} e^x\\
&=& {\G(a+x \d)\G(b+x \d)\over \G(c+x \d)} e^x \non\\
&=& \sum_{n=0}^\infty {\G(a+n)\G(b+n)\over \G(c+n)} {x^n\over n!}.\non
\eeqa
Using (\ref{Beq}), one obtains a Mellin-Barnes integral
representation\cite{AbS2}
\beq
{\G(a)\G(b)\over \G(c)}F(a,b;c;x)={1\over 2\pi i} \int_{-i\infty}^{i\infty}
{\G(a+s)\G(b+s)\G(-s)\over \G(c+s)} (-x)^s ds,
\eeq
where the contour separates the poles of $\G(a+s)\G(b+s)$ from those of
$\G(-s)$, $a,b,c$ are not negative integers, and $a-b$ is not an integer.

Another integral representation is found by recognizing that $C^{-1}$
times \newline
$\G(c-b)$ contains a beta function.  Thus, one finds
\beqa
\label{fint}
{\G(b)\G(c-b)\over \G(c)}F(a,b;c;x)&=&
{\G(b+x \d)\G(c-b)\over \G(c+x \d)} {\G(a+x \d)\over \G(a)}e^x \\
&=& \int_0^1 dt\, t^{b+x \d-1}(1-t)^{c-b-1}{\G(a+x \d)\over \G(a)}e^x \non
\eeqa
($\Re(c)>\Re(b)>0$).
This can be further simplified by expanding $e^x$ in a power series and
allowing $\G(a+x \d)$ to act on it.  This gives
\beq
{\G(a+x \d)\over \G(a)}e^x =\sum_{n=0}^\infty {\G(a+n)\over \G(a)\G(n+1)}
x^n.
\eeq
Using the gamma function inversion formula, $\G(z)\G(1-z)=\pi \csc\pi z$,
this gives
\beqa
{\G(a+x \d)\over \G(a)}e^x &=&\sum_{n=0}^\infty {\G(1-a)\over \G(1-a-n)\G(n+1)}
(-1)^n x^n \\
&=& (1-x)^{-a}. \non
\eeqa
Inserting this in (\ref{fint}), and letting $t^{x \d}$ act on $(1-x)^{-a}$,
one obtains the familiar integral representation\cite{AbS2}
\beq
{\G(b)\G(c-b)\over \G(c)}F(a,b;c;x)=
\int_0^1 dt\, t^{b-1}(1-t)^{c-b-1}(1-tx)^{-a}.
\eeq

The set of Kummer's 24 solutions of the hypergeometric equation can be
obtained by symmetry transformations which take the hypergeometric
equation into itself. Let $H(a,b,c)$ represent the hypergeometric
operator (\ref{hop}).
Since a quadratic
function of the coordinate multiplies the $\d^2$ term, the most general
shift of the derivative which produces a linear term of the correct form is
generated by $C_1= x^{-\a} (1-x)^{-\b}$,
\beq
\label{sim2}
\d\mapsto C_1\d C_1^{-1}=\d +{\a\over x} -{\b\over 1-x} .
\eeq
This produces a new Hamiltonian
\beqa
\H{a}&=& x(1-x) \d^2 +(c+2\a-(a+b+1+2\a+2\b)x)\d \\
&&+{\a^2-(1-c)\a \over x} +{\b^2+(a+b-c)\b\over
1-x}-(\a+\b+a)(\a+\b+b). \non
\eeqa
This is in hypergeometric form if the
undesired potential terms are cancelled by taking
\beqa
\a^2-(1-c)\a &=& 0 \\
\b^2 +(a+b-c)\b &=& 0. \nonumber
\eeqa
The solutions of the conditions are $\a=0,1-c$ and $\b=0,c-a-b$, and
these give three transformed hypergeometric functions.  (Note that these
are the exponents of the regular singular points at 0 and 1.)
The symmetry transformation is
\beq
C_1 H(a,b,c) C_1^{-1}= H(a+\a+\b,b+\a+\b,c+2\a).
\eeq
A solution of the
original hypergeometric equation is given in terms of the transformed one as
\beq
\psi=C_1^{-1}\ps{a}=x^{\a} (1-x)^{\b} F(\a+\b+a,\a+\b+b;c+2\a;x)
\eeq
For the four choices of $(\a,\b)$, one finds
\beqa
\ps1&=& F(a,b;c;x) \\
\ps2&=& x^{1-c} F(a-c+1,b-c+1;2-c;x) \non \\
\ps3&=& (1-x)^{c-a-b} F(c-a,c-b;c;x) \non \\
\ps4&=& x^{1-c} (1-x)^{c-a-b} F(1-b,1-a;2-c;x). \non
\eeqa
Inspecting the behavior at the regular singular point $x=0$, one finds that
$\ps1=\ps3$ and $\ps2=\ps4$.  For $c\ne 1$, $\ps1$ and $\ps2$ are
independent solutions.

In addition to similarity transformations, one can make point canonical
transformations.  The transformation $P_{1/x}$ takes
\beq
x\mapsto 1/x, \quad \d \mapsto -x^2 \d.
\eeq
This transforms the Hamiltonian to
\beq
\H{a}=-x[ x(1-x)\d^2 +(1-a-b-(2-c)x)\d +{ab\over x}].
\eeq
This is not in canonical form because of the term in $1/x$.  This can be
cancelled by making the similarity transformation $C_1$ (\ref{sim2}).
The transformed Hamiltonian is
\beqa
\label{Hb}
\H{b}&=&-x[ x(1-x)\d^2 +(1-a-b+2\a-(2\a+2\b+2-c)x)\d \\
&& \quad +{\a^2-\a(a+b)+ab\over x}+{\b^2+\b(a+b-c)\over 1-x} \non \\
&& \quad -(\a+\b)(1-c+\a+\b)]. \non
\eeqa
Taking $\a=a,b$ and $\b=0,c-a-b$ cancels the undesired potentials and
returns this to the hypergeometric form.  (Note that $\a$ is the exponent of
the regular singular point at $\infty$.)  One has
\beq
\label{c2}
C_1 P_{1/x} H(a,b,c) P_{1/x} C_1^{-1} = -x H(\a+\b,1-c+\a+\b,1-a-b+2\a).
\eeq
The overall factor of $x$ is
irrelevant since the equation is $\H{b}\ps{b}=0$, and it may be divided out.
The solutions of the original hypergeometric are
equation
\beq
\psi=P_{1/x}C_1^{-1}\ps{b}=
x^{-\a-\b}(x-1)^\b F(\a+\b,1-c+\a+\b;1-a-b+2\a;1/x)
\eeq
For the different choices of $(\a,\b)$, one finds
\beqa
\ps5&=& x^{-a} F(a,1-c+a;1+a-b; 1/x), \\
\ps6&=& x^{-b} F(b,1-c+b;1+b-a; 1/x), \non \\
\ps7&=& x^{b-c}(x-1)^{c-a-b} F(c-b,1-b;1+a-b; 1/x), \non \\
\ps8&=& x^{a-c}(x-1)^{c-a-b} F(c-a,1-a;1+b-a; 1/x). \non
\eeqa
Checking behavior at the regular singular point $x=\infty$, one finds
$\ps5=\ps7$ and $\ps6=\ps8$.

A second point canonical transformation $P_{1-x}$ also leads to new
solutions. This transforms
\beq
x\mapsto 1-x, \d\mapsto -\d.
\eeq
The transformed Hamiltonian is
\beq
\H{a}=x(1-x)\d^2+(1+a+b-c-(a+b+1)x)\d-ab.
\eeq
This is in hypergeometric form.  Applying the transformation $C_1$,
with $\a=0,c-a-b$, $\b=0,1-c$ (because $c$ in the original
hypergeometric equation has been replaced by $1+a+b-c$), one obtains
\beqa
\H{b}&=&x(1-x)\d^2+(1+a+b-c+2\a-(2\a+2\b+a+b+1)x) \d \non \\
&&\qquad -(\a+\b+a)(\a+\b+b).
\eeqa
One has
\beq
\label{c3}
C_1 P_{1-x} H(a,b,c) P_{1-x} C_1^{-1} = H(a+\a+\b,b+\a+\b,1+a+b-c+2\a),
\eeq
with $\a=0,c-a-b$, $\b=0,1-c$.
The solutions to the original hypergeometric equation are
\beq
\psi=P_{1-x}C_1^{-1}\ps{b}
=x^\b(1-x)^\a F(\a+\b+a,\a+\b+b;1+a+b-c+2\a;1-x).
\eeq
For the different choices of $(\a,\b)$, these are
\beqa
\ps9 &=& F(a,b;1+a+b-c;1-x), \\
\ps{10} &=& (1-x)^{c-a-b} F(c-b,c-a;1+c-a-b;1-x), \non \\
\ps{11} &=& x^{1-c} F(a-c+1,b-c+1;1+a+b-c;1-x), \non \\
\ps{12} &=& x^{1-c} (1-x)^{c-a-b} F(1-b,1-a;1+c-a-b;1-x). \non
\eeqa
Checking behavior at the regular singular point $x=1$, one finds
$\ps9=\ps{11}$ and $\ps{10}=\ps{12}$.

The previous two transformations can be composed to lead to further
solutions.  First, one can apply $C_1 P_{1/x} C_1 P_{1-x}$.  One must
remember that $C_1$ depends on two parameters $\a,\b$ that are not
reflected in the notation.  These parameters are determined
by the form of the hypergeometric function that $C_1$ acts upon, and are
in general different for each factor of $C_1$.  Using (\ref{c3}) and
(\ref{c2}), one finds
\beqa
C_1 P_{1/x} C_1 P_{1-x} H(a,b,c) P_{1-x} C_1^{-1} P_{1/x} C_1^{-1} &=& \\
&& \hspace{-3.5in} =C_1 P_{1/x} H(a+\a_1+\b_1,b+\a_1+\b_1,1+a+b-c+2\a_1)
P_{1/x} C_1^{-1} \non \\
&& \hspace{-3.5in} =-x
H(\a_2+\b_2,c-a-b-2\a_1+\a_2+\b_2,1-a-b-2\a_1-2\b_1+2\a_2), \non
\eeqa
where
\beqa
\a_1&=&0,\quad c-a-b,  \\
\b_1&=&0,\quad 1-c, \non \\
\a_2&=& \a_1+\b_1+a,\quad \a_1+\b_1+b, \non \\
\b_2&=& 0,\quad 1-c-2\b_1. \non
\eeqa
There are only four distinct forms of solution, and without loss of
generality one can take $(\a_1,\b_1)=(0,0)$.  This gives
\beqa
\ps{13} &=& (1-x)^{-a} F(a,c-b;1+a-b; {1\over 1-x}),  \\
\ps{14} &=& (1-x)^{-b} F(b,c-a;1+b-a; {1\over 1-x}), \non \\
\ps{15} &=& (-1)^{-a} x^{1-c} (x-1)^{c-a-1} F(1+a-c,1-b;1+a-b; {1\over 1-x}),
\non \\
\ps{16} &=& (-1)^{-b} x^{1-c} (x-1)^{c-b-1} F(1+b-c,1-a;1+b-a; {1\over
1-x}). \non
\eeqa

A second transformation is induced by $C_2=C_1 P_{1-x} C_1 P_{1/x}$.
Without loss of generality, one can take $(\a_2,\b_2)=(0,0)$, this
makes the symmetry transformation
\beq
C_2 H(a,b,c) C_2^{-1}
= (x-1) H(\a_1+\b_1,1-c+\a_1+\b_1,
1+a+b-c+2\b_1),
\eeq
where $\a_1=a,b$, $\b_1=0,c-a-b$.
The solutions of the hypergeometric equation are
\beqa
\ps{17} &=& x^{-a} F(a,a-c+1;a+b-c+1;1-{1\over x}),  \\
\ps{18} &=& x^{-b} F(b,b-c+1;a+b-c+1;1-{1\over x}), \non \\
\ps{19} &=& x^{b-c}(x-1)^{c-a-b} F(c-b,1-b;c-a-b+1;1-{1\over x}), \non \\
\ps{20} &=& x^{a-c}(x-1)^{c-a-b} F(c-a,1-a;c-a-b+1;1-{1\over x}), \non
\eeqa

Finally, the transformation induced by
$$
C_3=C_1 P_{1-x} C_1 P_{1/x} C_1 P_{1-x},
$$
taking $(\a_1,\b_1)=(0,0)$ and $(\a_2,\b_2)=(a,0)$, makes
gives the symmetry
\beq
C_3 H(a,b,c) C_3^{-1}=(x-1) H(a+\a_3+\b_3,c-b+\a_3+\b_3,c+2\a_3),
\eeq
where $\a_3=0,1-c$, $\b_3=0,b-a$. This gives the hypergeometric solutions
\beqa
\ps{21} &=& (1-x)^{-a} F(a,c-b;c;{x\over x-1}),  \\
\ps{22} &=& (-1)^{1-c}x^{1-c}(1-x)^{c-a-1} F(a-c+1,1-b;2-c;{x\over x-1}),
\non \\
\ps{23} &=& (1-x)^{-b} F(b,c-a;c;{x\over x-1}), \non \\
\ps{24} &=& (-1)^{1-c}x^{1-c}(1-x)^{c-b-1} F(b-c+1,1-a;2-c;{x\over x-1}). \non
\eeqa
This completes the construction of the 24 Kummer solutions of the
hypergeometric equation.

\def\vl{{\phantom{F}}}
\def\P{\protect}
\def\ft#1#2{{\textstyle{{#1}\over{#2}}}}
\def\arcsinh{\mathop{\rm arcsinh}\nolimits}
\def\cramp{\medmuskip = 2mu plus 1mu minus 2mu}
\def\crampest{\medmuskip = 1mu plus 1mu minus 1mu}
\def\uncramp{\medmuskip = 4mu plus 2mu minus 4mu}
\def\im{{\rm i}}

\section{Three-body Toda Quantum Mechanics}

As an application of canonical transformations to a more difficult problem,
consider the nonperiodic three-body Toda potential.  The Schr\"odinger
operator is
\beq
\label{toda}
H=-{1\over 2}(\d_1^2 +\d_2^2+\d_3^2) + e^{q_1-q_2}+e^{q_2-q_3}.
\eeq
This operator does not separate in any of the standard orthogonal
coordinate systems, yet integral representations for its eigenfunctions
can be constructed.  The n-body Toda potential, generalizing (\ref{toda})
in the obvious way, is physically interesting as an
exactly solvable nonlinear oscillator which may be obtained from the
Korteweg-deVries equation by an appropriate discretization\cite{Toda}.

     Two integral representations of the three-body Toda wavefunctions are
quoted in the review by Olshanetsky and Perelomov \cite{OlP}, and both can be
economically derived using canonical transformations.
The  result of Vinogradov and Takhtadjan \cite{ViT} was originally
found using methods from number theory.  A second form was
obtained in \cite{Bru} by constructing power series solutions for the
wavefunctions and then by simply stating the associated integral
representations.  The derivations of these integral representations
by canonical transformation are
analogous to the solutions of the hypergeometric equations, in that a
sequence of
transformations, one involving the Gamma function, is used to transform the
equation to a form where a solution is obvious by inspection.  An
alternative solution in which the Toda problem is transformed to a free
one with vanishing potential is discussed in \cite{ANPS}.

     The result of Vinogradov and Takhtadjan \cite{ViT} for the 3-body Toda
wavefunction is the most easily obtained and so we shall begin with it.
The 3-body Toda Hamiltonian (\ref{toda})
can be transformed into center-of-mass coordinates with separated
interaction potential by the transformation
$P_{{\rm\scriptscriptstyle C.O.M.}}$
\beq
\label{com}
\bega{lll}
q_1&\mapsto &\ \ \ft43 q_1 + \ft23 q_2 +\ft13 q_3,\hfill \\
q_2&\mapsto &- \ft23 q_1 + \ft23 q_2 +
\ft13 q_3,\hfill \\
q_3&\mapsto &-\ft23 q_1 -\ft43 q_2 + \ft13 q_3,\hfill
\enda
\qquad
\bega{lll}
\d_1 &\mapsto &\ \ \ft12 \d_1 +  \d_3\hfill \\
\d_2 &\mapsto &-\ft12 \d_1 + \ft12 \d_2 + \d_3\\
\d_3 &\mapsto & -\ft12 \d_2 + \d_3.
\enda
\eeq
The inverse transformation $P_{{\rm\scriptscriptstyle C.O.M.}}^{-1}$ is
\beqa
(q_1,q_2,q_3) &\mapsto& (\ft12 q_1 -\ft12 q_2, \ft12 q_2 - \ft12 q_3,
q_1+q_2+q_3) \\
(\d_1,\d_2,\d_3) &\mapsto& (\ft43 \d_1 -\ft23 \d_2 -\ft23 \d_3,
\ft23 \d_1+\ft23 \d_2- \ft43 \d_3, \ft13 \d_1+ \ft13 \d_2 +\ft13 \d_3). \non
\eeqa
The resulting Hamiltonian takes the form
\beq
\label{todab}
H^a=-{1\over 4}(\d_1^2+\d_2^2 -\d_1 \d_2) -{3\over 2} \d_3^2 + e^{2q_1} +
e^{2q_2}.
\eeq

The following sequence of transformations then
reduces the Hamiltonian to a form for which an eigenfunction can be found
by inspection.
\beq
\label{vttran}
\bega{lrll}
[{\rm VT1}]&\exp(-\ln 2 (\d_1+\d_2)):&\rlap{$\left\{\bega{lcr}
q_1 &\mapsto& q_1- \ln2,\\ q_2 &\mapsto& q_2- \ln2,
\enda \right.$}\hskip 3.75cm&\rlap{$\bega{lcr}
\d_1 &\mapsto& \d_1\\
\d_2 &\mapsto& \d_2
\enda$}\hskip 4.1cm\\[.3cm]
\P[{\rm VT2}]&\exp(-ik (q_1-q_2)):&\rlap{$\left\{\bega{lcr}
q_1&\mapsto& q_1,\\
q_2&\mapsto& q_2,
\enda
\right.$}\hskip 3.75cm&\rlap{$\bega{lcr}
\d_1 &\mapsto& \d_1+ik\\
\d_2 &\mapsto& \d_2-ik
\enda$} \hskip 4.1cm\\[0.3cm]
\P[{\rm VT3}]&{\Gamma(-{\d_1+\d_2\over 2}) \over
\Gamma(-{\d_1+3ik \over 2})
\Gamma(-{\d_2-3ik \over 2}) }:&\rlap{$\left\{\bega{lcr}
e^{2q_1} &\mapsto& {\d_1+3ik\over \d_1+\d_2} e^{2q_1},\\[0.1cm]
e^{2q_2} &\mapsto& {\d_2-3ik\over \d_1+\d_2} e^{2q_2},
\enda
\right.$}\hskip 3.75cm&\rlap{$\bega{lcr}
\d_1 &\mapsto& \d_1\\[0.1cm]
\d_2 &\mapsto& \d_2
\enda$}\hskip 4.1cm
\enda
\eeq
Suppressing the free-particle kinetic term $-\ft32 \d_3^2$ describing
the motion  of the center of mass, the Hamiltonian transforms as follows:
\beq
\label{vtHtran}
\bega{rrl}
& H^a=& -{1\over 4}(\d_1^2+\d_2^2 -\d_1 \d_2) + e^{2q_1} +
e^{2q_2}\\[0.1cm]
\P[{\rm VT1}]&\mapsto& -{1\over 4}(\d_1^2+\d_2^2 -\d_1 \d_2
-e^{2q_1} -  e^{2q_2})\non \\[0.1cm]
\P[{\rm VT2}]&\mapsto& -{1\over 4}(\d_1^2+3ik\d_1+\d_2^2 -3ik\d_2
-\d_1 \d_2 -e^{2q_1} -  e^{2q_2})+{3k^2\over 4}\non \\[0.1cm]
\P[{\rm VT3}]&\mapsto&
-{1\over 4(\d_1+\d_2)}  [(\d_1+3ik)(\d_1^2 -e^{2q_1})
+(\d_2-3ik)(\d_2^2-e^{2q_2})]
 + {3 k^2\over 4}.\non
\enda
\eeq
Inspection shows that this final Hamiltonian has
${\cal B}_\nu(e^{q_1}){\cal C}_\nu(e^{q_2})$  as an eigenfunction where
${\cal B}_\nu$, ${\cal C}_\nu$ stand for modified Bessel functions, each
either $I_\nu$, $K_\nu$ or a linear combination of them.
The eigenvalue of this eigenfunction is $\ft14(-\nu^2+3k^2)$.

     The eigenfunction of the
Toda Hamiltonian (\ref{toda}) (with zero center-of-mass momentum)
corresponding to this eigenfunction is
\beqa
\psi_{k,\nu}(q_1,q_2)&=&N_{k,\nu}P^{-1}_{{\rm\scriptscriptstyle C.O.M.}}
\exp\big((\d_1+\d_2)\ln 2\big)
\exp\big(ik(q_1-q_2)\big)\\
&&\qquad\qquad {\Gamma(-{\d_1+3ik\over 2}) \Gamma(-{\d_2-3ik\over 2})
\over \Gamma(-{\d_1+\d_2\over 2}) }
{\cal B}_\nu(e^{q_1}){\cal C}_\nu(e^{q_2}).\non
\eeqa
Recognizing the product of Gamma functions as a Beta function, and using  a
familiar integral representation for the Beta function, one finds the
result
\beqa
\psi_{k,\nu}
&=&N_{k,\nu}\,P^{-1}_{{\rm\scriptscriptstyle C.O.M.}}
e^{\im k(q_1-q_2)}
\int_0^{\infty} dt\,t^{-1-\ft12(\d_2-3\im k)}(1+t)^{\ft12(\d_1+\d_2)}
{\cal B}_\nu(2e^{q_1}){\cal C}_\nu(2e^{q_2})\non \\
&&\hspace{-0cm}=N_{k,\nu}\,e^{\ft12\im k(q_1-2q_2+q_3)}\int_0^{\infty}
dt\, t^{-1+\ft32\im k} \\
&&\qquad\qquad {\cal B}_\nu(2\sqrt{(1+t)}e^{\ft12(q_1-q_2)})
{\cal C}_\nu(2\sqrt{(1+t^{-1})}e^{\ft12(q_2-q_3)}).\non
\eeqa
The physically interesting solution is the wavefunction which dies in
the classically forbidden region, and this is found by taking
${\cal B}_\nu$ and ${\cal C}_\nu$ to be $K_\nu$.
With the identifications $k=-\ft12\im(t-s)$, $\nu=-1+\ft12(3s+t)$,
this agrees with the result of Vinogradov and Takhtadjan, as
quoted in \cite{OlP}.

     The solution given by Bruschi {\it et al.}\ \cite{Bru} can similarly be
found by this approach. The key idea is to begin
by factoring out the asymptotic plane-wave behaviour.  Denote the
momentum of this asymptotic plane wave  by
$\lambda=(\lambda_1,\lambda_2,\lambda_3)$ in the coordinate  system where the
Toda Hamiltonian takes the form (\ref{toda}). Also let
$a=\lambda_1-\lambda_2,\ b=\lambda_2-\lambda_3$. The sequence of
transformations to a theory for which an eigenfunction can be recognized by
inspection is
\beq
\bega{rrl}
&\ H=& -\ft12(\d_1^2+\d_2^2+\d_3^2)+e^{q_1-q_2} +e^{q_2-q_3}\\[0.2cm]
e^{-\im\lambda\cdot q}:&\mapsto& -{1\over 2}\big((\d_1+\im\lambda_1)^2
 +(\d_2+\im\lambda_2)^2+(\d_3+\im\lambda_3)^2\big) \\
&&\quad +e^{q_1-q_2} +e^{q_2-q_3} \\[0.2cm]
P_{{\rm\scriptscriptstyle C.O.M.}}:&\mapsto& -{1\over 4}
 (\d_1^2+\d_2^2-\d_1 \d_2 + 2\im a  \d_1
 +2\im b \d_2 - 4e^{2q_1} -4 e^{2q_2}) \\
 &&\quad -{3\over 2}\d_3^2
-\im (\lambda_1+\lambda_2+\lambda_3)\d_3  +{1\over
2}(\lambda_1^2+\lambda_2^2+\lambda_3^2)  \non
\enda
\eeq
where $P_{{\rm\scriptscriptstyle C.O.M.}}$ as in (\ref{com}).
Suppressing the eigenvalue $\ft12(\lambda_1^2+\lambda_2^2+\lambda_3^2)$
and terms involving $\d_3$ which describe the center of mass motion,
the transformations continue as
\beq
\bega{rrl}
e^{-\ln 2(\d_1+\d_2)}: &\mapsto& -{1\over 4}
 (\d_1^2+\d_2^2-\d_1 \d_2 + 2\im a  \d_1
 +2\im b \d_2 - e^{2q_1} - e^{2q_2}) \\[0.2cm]
\rlap{$ {\Gamma(1+\im a+\ft12\d_1) \over \Gamma(-\im b-\ft12 \d_2)
\Gamma(1+\im(a+b)+\ft12\d_1+\ft12\d_2) }:$}\hskip 3cm&&\\[0.2cm]
&\mapsto&
-{1\over 4(2\im (a+b) +\d_1+\d_2)}
 \bigl[ (\d_1+2\im a)(\d_1^2+2\im (a+b) \d_1 - e^{2q_1}) \non \\[0.1cm]
&&\qquad\quad
+(\d_2+2\im b)(\d_2^2+2\im (a+b) \d_2 +e^{2q_2})\bigr]  \\[0.2cm]
e^{\im(a+b)(q_1+q_2)}:&\mapsto& -{1\over
4(\d_1+\d_2)}\bigl[(\d_1+ \im (a-b))(\d_1^2- e^{2q_1}+(a+b)^2)\\[0.1cm]
 &&\quad+(\d_2+ \im (b-a))(\d_2^2 +e^{2q_2}+(a+b)^2)\bigr]
\enda
\eeq

The final Hamiltonian
in this sequence has $K_{\im(a+b)}(e^{q_1})H^{(1)}_{\im(a+b)}(e^{q_2})$ as an
eigenfunction with eigenvalue $\ft12(\lambda_1^2+ \lambda_2^2+\lambda_3^2)$.
This eigenfunction will lead to a solution which is asymptotically
free in the classically allowed region with asymptotic momentum
$\lambda=(\lambda_1,\lambda_2,\lambda_3)$ and which dies in the
classically forbidden region.  Other solutions to the equation can
be obtained by using other products of Bessel functions.
This corresponds to the Toda
solution of Bruschi {\it et al.}  Assembling the full transformation, this
solution is
\beqa
\label{full}
\psi_\lambda(q)&=& N_\lambda\,e^{\im\lambda\cdot q}
P^{-1}_{{\rm\scriptscriptstyle C.O.M.}} e^{\ln 2(\d_1+\d_2)}
{\Gamma\big(- \im b-\ft12 \d_2\big)
\Gamma\big(1+\im(a+b)+\ft12 \d_1+\ft12 \d_2\big) \over
\Gamma\big(1+\im a+\ft12 \d_1)\big)}
\non \\
&& \qquad e^{-\im(a+b)(q_1+q_2)}
K_{\im(a+b)}(e^{q_1})H^{(1)}_{\im(a+b)}(e^{q_2}).
\eeqa
The product of Gamma functions is once again a Beta function and has the
integral representation (with parameters at the limits of convergence)
\beqa
{\Gamma\big(-\im b-\ft12 \d_2\big)
\Gamma\big(1+\im(a+b)+\ft12\d_1+\ft12\d_2\big) \over
\Gamma\big (1+\im a+\ft12\d_1\big)}&=&\\
&&\hspace{-3cm}= e^{-b\pi+\ft{\im}2\pi \d_2}\int_0^\infty
d\tilde z_3\, (1-\tilde z_3)^{-1-\im a  -\ft12 \d_1}
\tilde z_3^{-1-\im b  -\ft12 \d_2}.\non
\eeqa
The translation operator $\exp(\im \pi \d_2/2)$ acts to transform
$H^{(1)}_{\im(a+b)}(e^{q_2})$ into $K_{\im(a+b)}(e^{q_2})$ times
some constant factors.

Letting the $\exp[\ln2(\d_1+\d_2)]$ factor act, one finds
\beqa
\psi_\lambda(q)&=&{-2\im N_\lambda\over\pi}\,e^{\im\lambda\cdot q}
P^{-1}_{{\rm\scriptscriptstyle C.O.M.}}
e^{-\pi b} 2^{-\im 2(a+b)} \int_0^\infty
d\tilde z_3\, (1-\tilde z_3)^{-1-\im a  -\ft12 \d_1}
\tilde z_3^{-1-\im b  -\ft12 \d_2} \\
&& e^{-\im(a+b)(q_1+q_2)}e^{(a+b)\pi}
K_{\im(a+b)}(2e^{q_1})K_{\im(a+b)}(2e^{q_2}) . \non
\eeqa
The integral representation for the modified Bessel function
\beq
\int_0^\infty dx\, x^{\nu-1} \exp\bl \im\mu(x-{e^{2q}\over x})\br=
2e^{\nu q} e^{i\nu \pi/2} K_{-\nu}(2\mu e^q),
\eeq
($Im(\mu)>0,\ Im(e^{2q}\mu)<0$) for $\nu=-\im(a+b)$,  with $\mu=1$
and $e^q$ appropriately displaced into the complex plane for convergence,
can be used to replace the Bessel functions.  Acting with the
translation operators appearing in the Beta function,  the result is
\beqa
\psi_\lambda(q)&=&{-\im N_\lambda\over2\pi}e^{-b\pi}2^{-\im2(a+b)}
e^{\im\lambda\cdot q}
\int_0^\infty d\tilde z_3\, (1-\tilde z_3)^{-\im a -1}
\tilde z_3^{-\im b -1} \\
&&\hspace{-1cm}\int_0^\infty dz_1 \int_0^\infty dz_2\,
(z_1  z_2)^{-\im(a+b)-1} e^{\im(z_1+z_2)}\exp\Big(-\im{e^{q_1-q_2}\over
z_2(1-\tilde z_3)} -\im {e^{q_2-q_3}\over z_1 \tilde z_3}\Big).\non
\eeqa
Finally, absorbing the constant factors into the  normalization
and making the change of variables $\tilde z_3=z_3/(z_1 z_2)$, one reaches
the form of Bruschi {\it et al.}
\beqa
\psi_\lambda(q)&=&N'_\lambda\,e^{\im\lambda\cdot q}
\int\limits_0^\infty\int\limits_0^\infty\int\limits_0^\infty dz_1 dz_2 dz_3\,
(z_1 z_2-z_3)^{\im (\lambda_2-\lambda_1) -1}
z_3^{\im (\lambda_3-\lambda_2) -1} \\
&&\qquad\qquad\qquad e^{\im(z_1+z_2)}\exp\Big(-\im{z_1 \over z_1 z_2-
z_3}e^{q_1-q_2} -\im {z_2\over  z_3} e^{q_2-q_3}\Big).\non
\eeqa
Note that this corrects some typos in the formula quoted in \cite{OlP}, most
importantly concerning the range of integration.  Note also that this is
an eigenfunction of (\ref{toda}) and not of the Hamiltonian of the form
discussed in Section 12 of \cite{OlP}.

The power of the quantum canonical transformations is evident in the
simplicity of these derivations.

\section{2-d Generalized Hypergeometric Functions}

The example of the 3-body Toda problem suggests a nontrivial
generalization of the hypergeometric equation to two dimensions.
The center of mass form of the Toda problem (\ref{todab}) has the form
of two Bessel operators in $q_1$ and $q_2$ with a cross term $\d_1 \d_2$.
The Gamma function transformation [VT3] in (\ref{vttran}) removes the cross
term and
effectively separates the problem so that a product of two Bessel functions
is a solution.  In the resulting full solution,  the Gamma function
transformation acts as a convolution of the Bessel functions.
Since a Bessel function is essentially a special form of hypergeometric
function, it is natural to consider the function obtained by similarly
convolving two (generalized) hypergeometric functions.

The characteristic feature of generalized hypergeometric
functions is the Gamma function structure of their series expansions
\beqa
\vl_r F_s(a_1,\ldots,a_r;b_1,\ldots,b_s;x)
&=& \sum_{n=0}^\infty {\prod_{j=1}^r \Gamma(a_j+ n)
\over \prod_{j=1}^r \Gamma(a_j)}{\prod_{k=1}^s \Gamma(b_k)
\over
\prod_{k=1}^s \Gamma(b_k+n)} {x^n\over n!} \\
&=&{\prod_{j=1}^r \Gamma(a_j+
x\d) \over \prod_{j=1}^r \Gamma(a_j)} {\prod_{k=1}^s \Gamma(b_k)\over
\prod_{k=1}^s \Gamma(b_k+x\d)} e^x. \non
\eeqa
$\vl_r F_s$ is a solution of the  operator
\beq
H= [\prod_{k=1}^s (b_k+x\d)]\d-\prod_{j=1}^r  (a_j+ x\d) ,
\eeq
The canonical transformation
\beq
C={\prod_{k=1}^s \Gamma(b_k+x\d) \over \prod_{j=1}^r  \Gamma(a_j+ x\d)}
\eeq
transforms $H$ to
\beq
H^a=[\prod_{j=1}^r  (a_j+ x\d)](\d-1),
\eeq
which has $e^x$ as a solution.

In a two-dimensional form, the intent is to couple two hypergeometric
equations with a cross term.  Define
\beqa
A_1&=&\prod_{j_1=1}^{r_1}  (a^{(1)}_{j_1}+ x_1\d_1),\\
A_2&=&\prod_{j_2=1}^{r_2}  (a^{(2)}_{j_2}+ x_2\d_2).\non
\eeqa
A natural equation is
\beqa
\label{genH}
H&=&A_1 \d_1 + A_2 \d_2 -[(x_1 \d_1)^2 + (x_2 \d_2)^2 -x_1 \d_1 x_2 \d_2\\
&&\quad +\a(x_1 \d_1 -x_2 \d_2) +\b].\non
\eeqa
By only making transformations which depend on $x_1 \d_1$ and $x_2 \d_2$,
only the isolated $\d_1$ and $\d_2$ factors will transform.  The
transformation
\beq
\label{Gam}
C_1=\Gamma(x_1\d_1+x_2 \d_2)
\eeq
takes
\beq
\d_1\mapsto {1\over x_1 \d_1 +x_2 \d_2}\d_1,\quad
\d_2\mapsto {1\over x_1 \d_1 +x_2 \d_2}\d_2.
\eeq
When the factor $ x_1 \d_1 +x_2 \d_2$ is multiplied into the expression
in square brackets, the cross term is cancelled and a separated expression
remains.  The result is
\beqa
\label{genHa}
H^a &=&{1\over x_1 \d_1 +x_2 \d_2}\bl A_1 \d_1 -(b_1 + x_1\d_1)
(b_2+ x_1 \d_1) x_1 \d_1 \\
&&\quad + A_2 \d_2 -(-b_1 +x_2 \d_2)(-b_2 +x_2 \d_2) x_2 \d_2 \br, \non
\eeqa
where $\a=b_1+b_2,\ \b=b_1 b_2$.  This has a product of two generalized
hypergeometric functions as a solution, and the desired result
has been obtained.

An operator of the form occuring in the three-body Toda equation can be
obtained from (\ref{genH}) by a few simple transformations which indicate
other natural forms of this 2-dimensional hypergeometric equation:  A Fourier
transform leads to an operator where the isolated $\d_1$, $\d_2$
become $x_1$ and $x_2$.  A point transformation $x_1\mapsto
x_1^k$, $x_2\mapsto x_2^k$ has the mild effect $x_1 \d_1\mapsto k^{-1}
x_1 \d_1$, $x_2 \d_2 \mapsto k^{-1} x_2 \d_2$, so one can consider
powers of the isolated factors different than 1.  Finally taking $k=2$
and $A_1=1=A_2$, an exponential point transformation
$x_1\mapsto e^{x_1},\ x_2\mapsto e^{x_2}$ results in an operator of the
form that the three-body Toda equation takes after [VT2] in (\ref{vtHtran}).

A superficial difference in the treatment of Toda is that the
Bessel functions are eigenfunctions with nonzero eigenvalue of the
Bessel operators and not simply the zero-eigenvalue solutions, as
the hypergeometric functions are here.  This is possible because
the factors multiplying the Bessel operators conspire to cancel
the leading factor $(x_1 \d_1 + x_2 \d_2)^{-1}$ when they both act on
the same function.  Appropriate $A_1$ and $A_2$ would allow
a similar freedom in (\ref{genHa}).   Or, one could use further
Gamma function canonical transformations to cancel $A_1$ and $A_2$
and substitute operators of appropriate form.  Using this freedom can be
a powerful way to obtain more explicit forms of solutions to the equation
than found by simply using the immediate solutions in terms of generalized
hypergeometric functions.

As well, one can make the observation that
the ``eigenvalue'' is implicit in the definition of the hypergeometric
operators.  By changing the constant $\beta$ in (\ref{genH}), one
gets solutions of the remaining operator with different eigenvalue.
The Bessel operator itself cannot be put in hypergeometric form until
the eigenvalue is specified, and alone it is in this sense incomplete.

It is not difficult to further generalize (\ref{genH}) by multiplying
the term in brackets by a product of polynomials in $x_1 \d_1$,
$x_2 \d_2$.  The remaining terms must also be appropriately modified
so that the operator which results after the transformation (\ref{Gam})
has a separable product as a solution.  Defining
\beqa
B_1&=&\prod_{k_1=1}^{s_1}  (b^{(1)}_{k_1}+ x_1\d_1),\\
B_2&=&\prod_{k_2=1}^{s_2}  (b^{(2)}_{k_2}+ x_2\d_2),\non
\eeqa
the appropriate generalization is
\beqa
H&=&B_2 A_1 \d_1 + B_1 A_2 \d_2 -B_1 B_2[(x_1 \d_1)^2 +
(x_2 \d_2)^2 -x_1 \d_1 x_2 \d_2\\
&&\quad +\a(x_1 \d_1 -x_2 \d_2) +\b].\non
\eeqa
After making the transformation induced by (\ref{Gam}), this becomes
\beqa
H^a &=&{1\over x_1 \d_1 +x_2 \d_2}\bl B_2[ A_1 \d_1 -B_1 (b_1 + x_1\d_1)
(b_2+ x_1 \d_1) x_1 \d_1] \\
&&\quad + B_1[ A_2 \d_2 -B_2(-b_1 +x_2 \d_2)(-b_2 +x_2 \d_2) x_2 \d_2] \br,
\non
\eeqa

It is clear that the essential step in solving (\ref{genH}) is
the transformation $\Gamma(x_1 \d_1 +x_2 \d_2)$.  In the series
expansion of the final solution, if the powers of $x_1$ and $x_2$ are
indexed by $n_1$ and $n_2$, respectively, this produces a factor of
$1/\Gamma(n_1+n_2)$.  It is not hard to see that this kind of trick could
be fruitfully applied to uncouple other equations and to produce
coupled factors in the series expansions of other functions.

\section{Conclusion}

Quantum canonical transformations have been used to give a simple and
economical derivation of the integral representations and Kummer
solutions of the confluent hypergeometric and hypergeometric equations.
It is straightforward to extend this discussion to derive the
differentiation
formulae relating hypergeometric functions of different
indices.  The transformations to the specific forms of the equations for
the many classical special functions are also easily handled
with canonical transformations.  The quantum canonical transformations
are seen to provide a useful systematization of techniques for
solving differential equations.

Two integral representations for the physically interesting solutions of
the non-periodic three-body Toda equation were also found
using quantum canonical transformations.
These motivated a nontrivial two-dimensional generalized hypergeometric
equation of which the non-periodic three-body Toda equation is
a special case.  The methods described here can be applied
to solve other nonseparable partial differential equations which do not
separate in the standard orthogonal coordinate systems.

\end{document}